\documentclass[aps,pre,10pt,twocolumn,floats,floatfix,showpacs,showkeys,superscriptaddress]{revtex4}
\usepackage{graphicx,amssymb,amsmath}
\usepackage[latin1]{inputenc}
\usepackage{color}

\definecolor{red}{rgb}{1,0,0}
\definecolor{green}{rgb}{0,1,0}
\definecolor{blue}{rgb}{0,0,1}
\definecolor{darkmagenta}{rgb}{.5,0,.5}

\begin{document}
\title{General analytical solutions for DC/AC circuit network analysis}
\author{Nicol{\'a}s Rubido}
\affiliation{Institute for Complex Systems and Mathematical Biology, University of Aberdeen,
	     King's College, AB24 3UE Aberdeen, U.K.}
\affiliation{Instituto de F\'{i}sica, Facultad de Ciencias, Universidad de la Rep\'{u}blica,
	     Igu\'{a} 4225, Montevideo, 11200, Uruguay}
\author{Celso Grebogi}
\affiliation{Institute for Complex Systems and Mathematical Biology, University of Aberdeen,
	     King's College, AB24 3UE Aberdeen, U.K.}
\author{Murilo S. Baptista}
\affiliation{Institute for Complex Systems and Mathematical Biology, University of Aberdeen,
	     King's College, AB24 3UE Aberdeen, U.K.}
\date{\today}
\begin{abstract}
In this work, we present novel general analytical solutions for the currents that are
developed in the edges of network-like circuits when some nodes of the network act as
sources/sinks of DC or AC current. We assume that Ohm's law is valid at every edge and
that charge at every node is conserved (with the exception of the source/sink nodes). The
resistive, capacitive, and/or inductive properties of the lines in the circuit define a
complex network structure with given impedances for each edge. Our solution for the currents
at each edge is derived in terms of the eigenvalues and eigenvectors of the Laplacian matrix
of the network defined from the impedances. This derivation also allows us to compute the
equivalent impedance between any two nodes of the circuit and relate it to currents in a
closed circuit which has a single voltage generator instead of many input/output source/sink
nodes. Contrary to solving Kirchhoff's equations, our derivation allows to easily calculate
the redistribution of currents that occurs when the location of sources and sinks changes
within the network. Finally, we show that our solutions are identical to the ones found
from Circuit Theory node analysis.
\end{abstract}
%
\pacs{41.20.-q,89.75.Hc,45.30.+s,95.75.Pq}
\keywords{Circuit networks, Laplacian matrix, Ohm's law, Kirchhoff's laws, Resistor networks, Impedances.}
\maketitle
\section{Introduction}
The currents in each edge of an electrical circuit, which is composed of
linear elements (i.e., resistance, capacitance, and inductance) and where conservation of
charge at each node is granted, are generally found by solving Kirchhoff's equations
\cite{Kirchhoff}. In particular, for resistor networks, the solution for the currents at
each edge is related to random walks in graphs \cite{FanChung1997}, first-passage times
\cite{Randall2006}, finding shortest-paths and community structures on weighted networks
\cite{Newman2004}, and network topology spectral characteristics \cite{Rubido2013a,
Rubido2013b}. Though the relationship between currents and voltage differences in network
circuits with linear elements follows Ohm's law, their modelling capability is enormous.
For example, it is used to model fractures in materials \cite{Batrouni1998}, biologically
inspired transport networks \cite{Katifori2010}, airplane traffic networks
\cite{Bocaletti2013}, robot path planing \cite{Zuojun2006}, queueing systems
\cite{Haenggi2002}, etc.

In practice, resistor networks are used in various electronic designs, such as current or
voltage dividers, current amplifiers, digital to analogue converters, etc. These devices are
usually inexpensive, relatively easy to manufacture, and require little precision on the
constituents. In order to solve the voltages across these networks, two methods are broadly
used: nodal analysis and mesh analysis \cite{Kirchhoff}. In the former, nodes are labelled
arbitrarily and voltages are set by using the Kirchhoff's current equations of the system.
In the later, loops are defined with an assigned current which do not contain any inner
loop, then the Kirchhoff's voltage equations are solved. These constitute classic techniques
of Circuit Theory.

\begin{figure}[htbp]
 \begin{center}
  \textbf{(a)}
  \includegraphics[width=1.0\columnwidth]{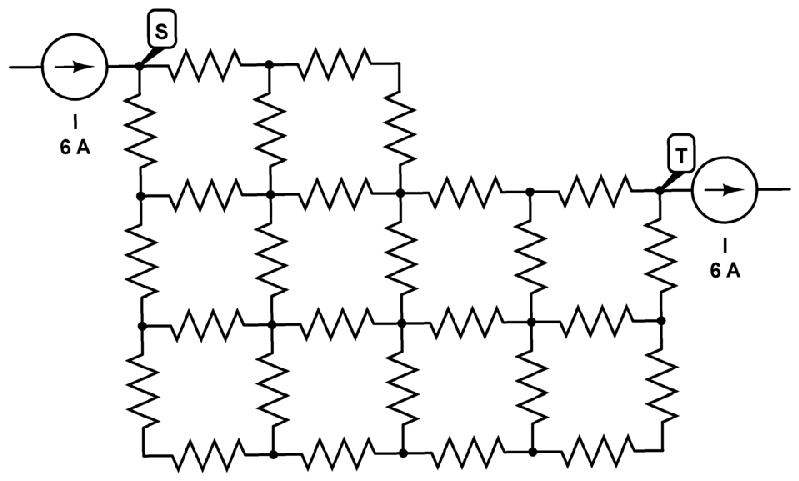} \\
  \textbf{(b)}
  \includegraphics[width=1.0\columnwidth]{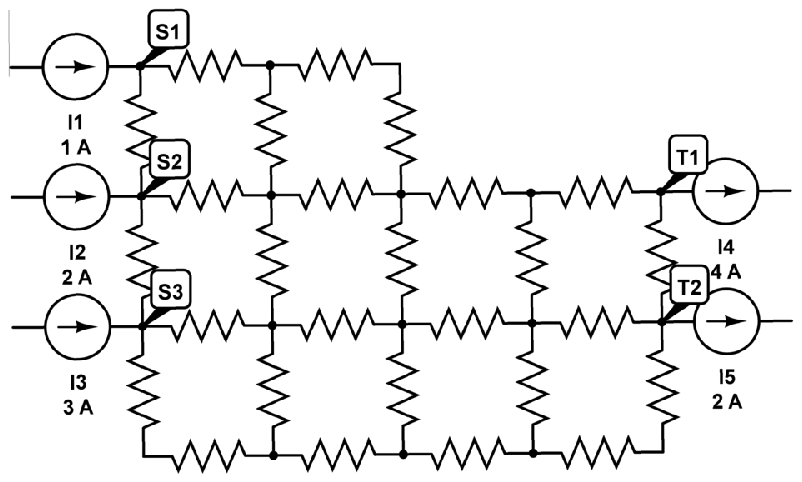}
 \end{center}
  \caption{Panel {\bf (a)} shows a schematic representation of a resistor network with a
  single input current of $I = 6\,A$ at a source node $S$ and a single output current of
  $I = 6\,A$ at a sink node $T$. Panel {\bf (b)} shows the same resistor network but with
  multiple inputs $I_1,I_2,I_3$ (nodes $S_1,S_2,S_3$) and outputs $I_4,I_5$ (nodes $T_1,T_2$)
  which add to the same inflow/outflow magnitudes than in panel {\bf (a)}. Changing the
  system from panel {\bf (a)} to panel {\bf (b)}, or vice-versa, generates a global
  redistribution of currents.}
 \label{fig_1}
\end{figure}

However, nodal and mesh methods (or even transfer function methods \cite{Ali2010}) become
inefficient to recalculate the voltage drops across the network if the location of inputs
and outputs changes constantly, e.g., if the cathode and/or anode of a voltage generator
are moved from one node of the network to another. This switching situation is common in
the modelling of the modern power-grid as an impedance network circuit or in general
supply-demand networks \cite{Rubido2013a,Rubido2013b,Batrouni1998,Katifori2010}. An example
of this case is shown in Fig.~\ref{fig_1} for a resistor network with a single source-sink
nodes. Another redistribution of currents, which is also poorly accounted by these methods,
happens if a single source node and single sink node are decentralized for multiple source
and/or multiple sink nodes that preserve the initial input and output magnitudes
\cite{Batrouni1998,Renatas2013}. In any case, either of the classical Circuit Theory
methods requires to be applied for each configuration of the sources and sinks in order
to find the currents at every edge of the network.

In this work, we present novel general analytical solutions for current conservative DC/AC
circuit networks with resistive, capacitive, and/or inductive edge characteristics. The
novelty comes from expressing the currents and voltage drops in terms of the eigenvalues
and eigenvectors of the admittance (namely, the inverse of the edge impedance) Laplacian
matrix of the circuit network.

In order to derive our novel solutions we assume that the impedance values at every edge
and the location of the source/sink nodes are known. Our solutions give the exact DC/AC
currents that each edge of the circuit holds and are identical in magnitude to the ones
found from nodal Circuit Theory analysis. The practicality of our solutions comes from,
allowing to compute the equivalent impedance between any two nodes of the network directly
\cite{Cserti2000,Wu2004,Jason2012} and allowing to easily calculate the redistribution of
currents that happens when the location of sources and sinks is changed within the network
(such as in the example of Fig.~\ref{fig_1}). The scientific interest of our solutions
comes from, establishing a clear relationship between the currents and voltages in DC/AC
circuits with the topology invariants of the network, namely, its eigenvalues and
eigenvectors.

The approach we develop provides new analytical insight into the transmission flow problem
and exhibits different features than other available solutions. Moreover, it provides a new
tool to achieve the voltage/current solutions and to analyse resonant behaviour in linear
circuits \cite{Silva2013}. As a practical application, we relate these solutions to closed
circuits where a voltage generator is present (instead of having open sources/sinks that
feed current to the network) and solve a simple network where we can compare our solutions
to the ones provided by solving directly Kirchhoff's nodal equations.

\section{DC/AC circuit networks}
 \subsection{The mathematical model}
The model we solve corresponds to a conservative circuit network with known input/output
net currents and obeys Ohm's and Kirchhoff's law for conservation of charge. We assume that
the input (output) net current $\sum_k I_{sk}$ ($\sum_k I_{tk}$) at the source (sink) node
$s$ ($t$), its frequency $\omega$ (with $\omega = 0$ for DC currents and $\omega>0$ for AC
currents), and their phases are known. The extension to various input or output nodes is
done in Appendix \ref{Anex_3}.

Ohm's law linearly relates the current at an edge of the circuit with the voltage difference
between the nodes that the edge connects. Specifically,
\begin{equation}
  I_{kl}^{(s,t)} = \frac{ V_k^{(s,t)} - V_l^{(s,t)} }{ Z_{kl} } = \frac{ \Delta
   V_{kl}^{(s,t)} }{ Z_{kl} }\,,
 \label{eq_Ohm}
\end{equation}
where $I_{kl}^{(s,t)}$ is the current passing from node $k$ to node $l$ given a current
source located at node $s$ and a sink node located at node $t$, $\Delta V_{kl}^{(s,t)}$
is the voltage difference, and $Z_{kl} = Z_{lk}$ is the impedance of the symmetric edge.
$Z_{kl}$ depends on the edge's resistive, capacitive, and/or inductive properties, and the
network topological properties of the circuit.

The variables in Eq.~(\ref{eq_Ohm}) are complex numbers in the case of AC input/output
currents and real numbers for DC currents. In general, a resonance in the $kl$-edge appears
for a minimum of the impedance, namely, when the input/output frequency $\omega$ is tuned
to a frequency related to the natural frequency of the edge line. For example, in the case
that the edge is modelled by a series $RLC$ circuit, the impedance of the edge is
\begin{equation}
  Z_{kl} = R_{kl} + j\,\omega\,L_{kl} \left[ 1 - \left( \frac{{\omega_{kl}^{(LC)}}}{\omega}
   \right)^2 \right] = Z_{lk}\,,
 \label{eq_Z_RLC}
\end{equation}
where $\omega_{kl}^{(LC)} = \sqrt{1/L_{kl} C_{kl}}$ is the natural frequency of the edge,
$\gamma_{kl}^{(RL)} = R_{kl}/L_{kl}$ is the dissipation of the edge, $L_{kl}$ is the edge's
inductance, $C_{kl}$ is the edge's capacitance, $R_{kl}$ is the edge's resistance, and $j =
\sqrt{-1}$. In this case, a resonance in the $kl$-edge appears if $\omega = \omega_{kl}^{
(LC)}$. Consequently, our solutions are valid as long as the input/output frequency $\omega$
is different from any of the $M$ resonant frequencies associated to the $M\in[N,N(N-1)/2]$
edges of the network.

The topological properties of the network are also included in the value of the impedance
of each edge. For instance, the impedance between two nodes that are not connected is
assumed to be $\left| Z_{kl} \right| = \infty$ because $R_{kl} = \infty$ for a non-existing
line. Otherwise, the impedance of two connected nodes is $\left| Z_{kl} \right| < \infty$.
Hence, the inverse of the impedance defines a complex valued adjacency matrix, i.e., it
gives which are the edges connecting the nodes in the circuit and their weights.

Kirchhoff's law of conservation of charge states that current is conserved at every node
in the circuit. In other words, the inflow equals the outflow at a node, with the exception
of the source and sink nodes (the extension to multiple sources and sinks is detailed in
Appendix \ref{Anex_3}). Specifically,
\begin{equation}
  \sum_{l = 1}^N I_{kl}^{(s,t)} = \delta_{ks}\,F^{in} - \delta_{kt}\,F^{out}\,,
 \label{eq_netflow}
\end{equation}
where $F^{in}$ ($F^{out}$) is the complex valued net current inflow (outflow) and
$\delta_{ks}$ ($\delta_{kt}$) is the Kronecker delta function. By assuming local
conservation of charge at every node of the circuit, we get that the total flow in the
network needs to be null, namely, global conservation of charge holds: $\sum_k \sum_l
I_{kl}^{(s,t)} = 0$, which is fulfilled only if $F^{in} = F^{out} = F$.

Using Eqs.~(\ref{eq_Ohm}) and (\ref{eq_netflow}), the model equations are
\begin{equation}
  V_k^{(s,t)} \sum_{l = 1}^N \frac{1}{ Z_{kl} } - \sum_{l = 1}^N\frac{ V_l^{(s,t)} }{ Z_{kl} }
   = F\,\left(\delta_{ks} - \delta_{kt}\right)\,,
\end{equation}
which are expressed in matrix form as
\begin{equation}
  \mathbf{G}\,\vec{V}^{(s,t)} = \vec{F}^{(s,t)}\,,
 \label{eq_model}
\end{equation}
where $\mathbf{G}$ is the weighted admittance Laplacian matrix of the network [with complex
entries given by $G_{kl} = \delta_{kl}(\sum_{m=1}^N 1/Z_{km}) + (\delta_{kl} - 1)/Z_{kl}$],
$\vec{F}^{(s,t)}$ is the inflow/outflow vector (with non-zero entries only at node $s$, $F$,
and at node $t$, $-F$), and $\vec{V}^{(s,t)}$ is the voltage potential at each node of the
network. Properties of the weighted Laplacian matrix $\mathbf{G}$ are dealt in Appendix
\ref{Anex_1}.

 \subsection{The analytical solutions}
We are deriving two main analytical results in this work. An expression for the DC/AC
currents flowing through each edge of the network (as function of the location of the
source and sink nodes and the net inflow magnitude) and the equivalent impedance between
any two nodes of the network. Both results are expressed in terms of the eigenvalues and
eigenvectors of the Laplacian matrix $\mathbf{G}$ of Eq.~(\ref{eq_model}), and are found
from the inversion of $\mathbf{G}$ (see Appendix \ref{Anex_2} for details on the inversion
of $\mathbf{G}$).

We find that after the inversion of $\mathbf{G}$ the voltage difference between any two
nodes of the circuit is given by
\begin{equation}
  \Delta V_{kl}^{(s,t)} = F \sum_{n = 1}^{N-1} \frac{\left( \left[\vec{v}_n\right]_k -
   \left[\vec{v}_n\right]_l \right)}{\lambda_n(\mathbf{G})} \left( \left[\vec{v}_n
    \right]_s^\star - \left[\vec{v}_n\right]_t^\star \right)\,,
 \label{eq_volt_diff}
\end{equation}
where $\lambda_n(\mathbf{G})$ is the complex $n$-th eigenvalue of Laplacian $\mathbf{G}$
(with $n = 0,1,\ldots,N-1$ and $\lambda_0(\mathbf{G}) = 0$ \cite{FanChung1997}) and
$\left[\vec{v}_n\right]_k$ is the corresponding $n$-th eigenvector $k$ coordinate (with
$k = 1,\ldots,N$). With the exception of $F$ (assuming the phase difference between the
net input and output flows is null, which guarantees global charge conservation), the
remaining quantities are complex numbers, hence, they have an amplitude and a phase,
and the $\star$ indicates complex conjugation. Thus,
\begin{equation}
  \Delta V_{kl}^{(s,t)} = F \sum_{n = 1}^{N-1} \left[ \alpha_{kl}^{(st)}(n) + j\,
   \beta_{kl}^{(st)}(n) \right]\,\frac{ e^{j\,\phi_{kl}^{(st)}} }{ \lambda_n(\mathbf{G}) }\,,
 \label{eq_volt_diff_comp}
\end{equation}
where $\alpha_{kl}^{(st)}(n)$ [$\beta_{kl}^{(st)}(n)$] is the real [imaginary] part of the
product $\left( \left[\vec{v}_n\right]_k - \left[\vec{v}_n\right]_l \right) \left( \left[
\vec{v}_n\right]_s^\star - \left[\vec{v}_n\right]_t^\star \right)$ and the phases
$\phi_{kl}^{(st)}$ are
\begin{equation}
  \phi_{kl}^{(st)} = \frac{ - \alpha_{kl}^{(st)}(n)\,\lambda_n(\mathbf{G}_I) + \beta_{kl}^{
   (st)}(n)\,\lambda_n(\mathbf{G}_R) }{ \alpha_{kl}^{(st)}(n)\,\lambda_n(\mathbf{G}_R) +
    \beta_{kl}^{(st)}(n)\,\lambda_n(\mathbf{G}_I) }\,,
 \label{eq_volt_diff_phase}
\end{equation}
with $\lambda_n(\mathbf{G}_R)$ [$\lambda_n(\mathbf{G}_I)$] being the $n$-th eigenvalue of
the real [imaginary] part of the Laplacian matrix $\mathbf{G}$ (details on the properties
of these eigenvalues are provided in Appendix \ref{Anex_1}).

Our first main analytical result is the current passing through the $kl$-edge, namely,
\begin{equation}
  I_{kl}^{(s,t)} = \frac{F}{Z_{kl}} \sum_{n = 1}^{N-1} \left[ \alpha_{kl}^{(st)}(n) +
   j\,\beta_{kl}^{(st)}(n) \right]\,\frac{ e^{j\,\phi_{kl}^{(st)}} }{ \lambda_n(\mathbf{G}) }\,,
 \label{eq_current}
\end{equation}
where $Z_{kl} = \left| Z_{kl} \right|\,e^{j\,\varphi_{kl}}$, $\varphi_{kl}$ being the
impedance phase value. This phase also corresponds to the phase difference between the
voltage drop $\Delta V_{kl}^{(s,t)}$ between nodes $k$ and $l$ and the current $I_{kl}^{
(s,t)}$ at the $kl$-edge.

Setting the source at node $k$ and the sink at node $l$, Eq.~(\ref{eq_volt_diff}) results
in
\begin{equation}
  \Delta V_{kl}^{(k,l)} = F \sum_{n = 1}^{N-1} \frac{ \left|\, \left[\vec{v}_n\right]_k -
   \left[\vec{v}_n\right]_l\, \right|^2}{ \lambda_n(\mathbf{G}) }\,,
\end{equation}
and the second main analytical result is derived, i.e.,
\begin{equation}
  Z_{kl}^{(eq)} = \sum_{n = 1}^{N-1} \frac{ \left|\, \left[\vec{v}_n\right]_k -
   \left[\vec{v}_n\right]_l\, \right|^2}{ \lambda_n(\mathbf{G}) }\,.
 \label{eq_equiv_impedance}
\end{equation}
$Z_{kl}^{(eq)}$ is the effective weight that all edges linking node $k$ with $l$ weigh
\cite{Wu2004,Rubido2013a,Rubido2013b}, namely, the equivalent impedance. Its value is
identical to the ones obtained by using Green functions \cite{Cserti2000} or Circuit Theory
analysis \cite{Kirchhoff}. For example, if all series and parallel impedances between nodes
$k$ and $l$ are added, then the final value is equal to the one that is found from
Eq.~(\ref{eq_equiv_impedance}).

 \subsection{Practical examples}
Here we show how to relate the solution found so far [Eq.~(\ref{eq_current})] for the
currents in the edges of a network-like circuit for given input/output nodes with the
currents in the edges of a closed circuit with a single voltage generator. As it is known,
solutions for closed linear circuits with given boundary voltage values correspond to
finding solutions of the Laplace equation without sources, which are widely known. Hence,
finding a relationship between Eq.~(\ref{eq_current}) with the closed circuit currents, in
our case, is equivalent to finding a relationship between topological invariant properties
of the network structure (impedances) with Laplace boundary problem solutions.

\begin{figure}[htbp]
 \begin{center}
  \textbf{(a)}
  \includegraphics[width=1.0\columnwidth]{network_flow_circuit_single_st.eps} \\
  \textbf{(b)}
  \includegraphics[width=1.0\columnwidth]{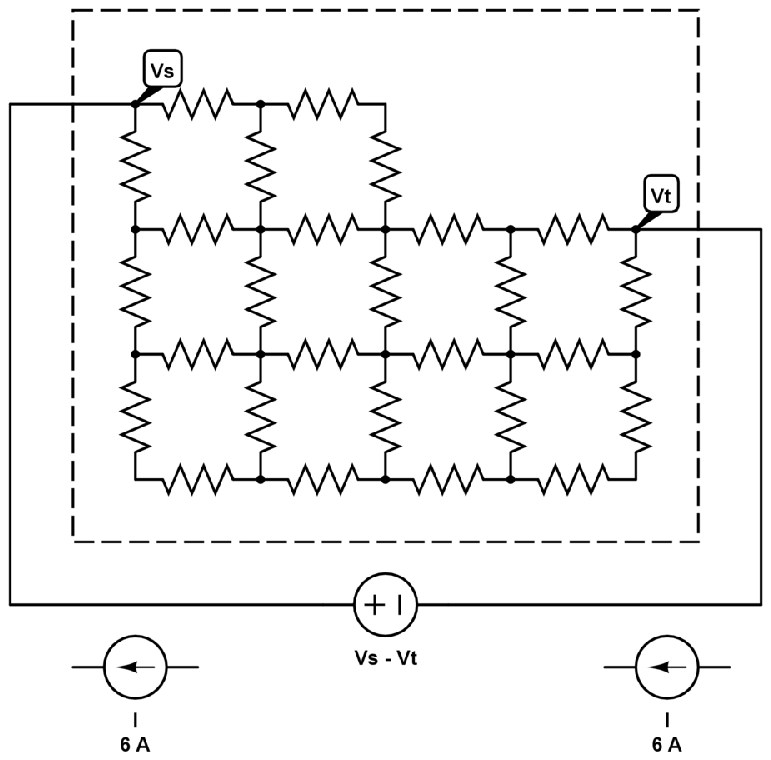}
 \end{center}
  \caption{Panel {\bf (a)} shows a schematic representation of a resistor network with a
  single input current of $I = 6\,A$ at a source node $S$ and a single output current of
  $I = 6\,A$ at a sink node $T$. Panel {\bf (b)} shows the same resistor network but with
  a voltage generator and a resistive line connecting nodes $S$ and $T$ such that the inflow
  (outflow) at node $S$ ($T$) equals $I$.}
 \label{fig_2}
\end{figure}

For a single source-sink problem [like the one depicted in Fig.~\ref{fig_2}{\bf (a)}]
to be transformed to a Laplace problem [Fig.~\ref{fig_2}{\bf (b)}], the source and sink
nodes ($S$ and $T$) are plugged into a voltage generator. The known data from the
source-sink problem is the fixed input net current $F$, which the generator will need to
provide to the network circuit between nodes $S$ and $T$. Hence, the voltage requirement
for the generator to supply is: $\epsilon_{st} = V_s - V_t = \rho_{st}\,F$, where
$\rho_{st}$ is the equivalent resistance between the nodes $S$ and $T$ of the known network
circuit structure [dashed square in Fig.~\ref{fig_2}{\bf (b)}]. The equivalent resistance
is found from \cite{Randall2006,Rubido2013a,Rubido2013b}
\begin{equation}
  \rho_{st} = \sum_{n = 1}^{N-1} \frac{1}{\lambda_n}\left( \left[\vec{v}_n\right]_s -
   \left[\vec{v}_n\right]_t \right)^2\,,
 \label{eq_equiv_resist}
\end{equation}
where $\lambda_n$ ($\vec{v}_n$) is the real $n$-th eigenvalue (eigenvector) of the resistor
network Laplacian matrix (which is also positively defined because its entries are all real
valued). Equation~(\ref{eq_equiv_resist}) is the resistive version of
Eq.~(\ref{eq_equiv_impedance}). Then, the corresponding identical Laplace problem to the
single source-sink pair of nodes is solved for the border conditions corresponding to a
voltage generator supplying a constant voltage difference $\epsilon_{st} $ between nodes
$S$ and $T$ of magnitude
\begin{equation}
  \epsilon_{st} = \rho_{st}\left(\mathcal{G}\right)\,F\,,
 \label{eq_voltgen_single_st}
\end{equation}
where $\mathcal{G} = \{\mathcal{V},\,\mathcal{E}\}$ is the node and edge set which define
the circuit network. Equation~(\ref{eq_voltgen_single_st}) establishes a direct relationship
between solving Laplace problems in circuits with transportation problems. Consequently,
this relationship increases the importance of our voltage solution in terms of the Laplacian
matrix eigenvalues and eigenvectors.

As another practical example, we compare our solutions for a square-like resistor network
with equal edges with the solutions found from linear Circuit Theory analysis. We index the
nodes in the square as $s,\,a,\,t,\,b$ in clockwise direction and the edge's resistances as
$R_{sa} = R_{at} = R_{sb} = R_{bt} = R$. Setting an input (output) source (sink) at node $s$
($t$) with magnitude $F_s^{(st)} = I$ ($F_t^{(st)} = -I$), nodal analysis gives the
following results for the edge currents that this circuit system has: $I_{sa} = I_{sb} =
I/2$ and $I_{ta} = I_{tb} = -I/2$. Also, the resultant equivalent resistance between nodes
$s$ and $t$ for the square is given by
\begin{equation}
  \rho_{st} = \left( \frac{1}{R_{sa} + R_{at}} + \frac{1}{R_{sb} + R_{bt}} \right)^{-1} = R\,.
 \label{eq_square_equivres}
\end{equation}

In our framework, we transform the resistor network into a topological problem, i.e., we
analyse the Laplacian matrix $\mathbf{G}$ of the network. Applying Ohm's and Kirchhoff's
laws, a square-like resistor network has the following conductance matrix $\mathbf{G}$
[see Eq.~(\ref{eq_model})]
\begin{equation}
 \mathbf{G} = \frac{1}{R} \left(\begin{array}{cccc}
				  2 & -1 & -1 & 0 \\
				  -1 & 2 & 0 & -1 \\
				  -1 & 0 & 2 & -1 \\
				  0 & -1 & -1 & 2
				 \end{array}\right) = \frac{1}{R}\mathbf{L}\,,
 \label{eq_prob_Lap}
\end{equation}
where the first column/row corresponds to node $s$, then node $a$, node $b$, and finally node
$t$. The eigenvalues of $\mathbf{G}$ are $\lambda_0 = 0$, $\lambda_1 = \lambda_2 = 2/R$, and
$\lambda_3 = 4/R$, and the eigenvectors are $\vec{v}_0 = \frac{1}{2}\left(1,1,1,1\right)^T$,
$\vec{v}_1 = \frac{1}{\sqrt{2}}\left(0,-1,1,0\right)^T$, $\vec{v}_2 = \frac{1}{\sqrt{2}}
\left(1,0,0,-1\right)^T$, and $\vec{v}_3 = \frac{1}{2}\left(-1,1,1,-1\right)^T$. Thus, using
Eq.~(\ref{eq_volt_diff}) for the edge current between nodes $s$ and $a$, we have
$$ I_{sa}^{(st)} = \frac{\Delta V_{sa}^{(st)}}{R} = \sum_{n = 1}^3 \left( \left[
\vec{v}_n\right]_s - \left[\vec{v}_n\right]_a \right) \frac{I}{R\,\lambda_n} \left(
\left[\vec{v}_n\right]_s - \left[\vec{v}_n\right]_t \right), $$
that results in
\begin{equation}
  I_{sa}^{(st)} = I \left( \left[\vec{v}_2\right]_s - \left[\vec{v}_2\right]_a \right)
   \frac{1}{2} \left( \left[\vec{v}_2\right]_s - \left[\vec{v}_2\right]_t \right) =
    \frac{I}{2},
 \label{eq_current_sol}
\end{equation}
where the other eigenvector modes in the sum are cancelled or have null coordinates.
Similarly, the remaining edge currents are calculated and found identical to the ones
from nodal analysis. Moreover, the equivalent resistance we calculate for nodes $s$ and
$t$ using Eq.~(\ref{eq_equiv_resist}) is
\begin{equation}
 \rho_{st} = \frac{\left( \left[\vec{v}_2\right]_s - \left[\vec{v}_2\right]_t \right)^2}{2/R}
  = R\,,
\end{equation}
which again, is identical to the one in Eq.~(\ref{eq_square_equivres}).

We can extend this problem easily for the case where the square circuit has equal impedances
$Z = R + j\,\omega\,L\,[1 - (\omega_0^2/\omega)^2]$ in every edge, where $\omega_0^2 = 1/LC$
is the characteristic frequency of each edge and $\omega$ is the input frequency ($F_s^{(st)
}(T) = I\,e^{j\,\omega\,T} = F_t^{(st)}(T)$, for every time $T$). Then, the admittance
Laplacian matrix entries from Eq.~(\ref{eq_model}) are given by
\begin{equation}
  G_{kl} = \delta_{kl}\left( \sum_{m=1}^4 \frac{1}{Z_{km}} \right) - \frac{(\delta_{kl} - 1)
            }{ Z_{kl} }\,.
 \label{eq_prob_Lap_complex}
\end{equation}
The inverse of the impedance (admittance) is given by
\begin{equation}
 \frac{1}{Z_{kl}} = A_{kl}\frac{e^{-j\,\varphi}}{\left| Z \right|}\,,
\end{equation}
where $A_{kl} = 1$ if node $k$ is connected to node $l$, $A_{kl} = 0$ otherwise, and $\tan(
\varphi) = L\,(\omega^2 - \omega_0^2)/R\omega$. The resultant admittance Laplacian matrix
in this case is
\begin{equation}
 \mathbf{G} = \mathbf{G}_R + j \mathbf{G}_I = \frac{e^{-j\,\varphi}}{\left| Z \right|}
  \mathbf{L} = \frac{\mathbf{L}}{Z}\,,
\end{equation}
with $\mathbf{G}_R$ ($\mathbf{G}_I$) being the real (imaginary) part of the entries in
Eq.~(\ref{eq_prob_Lap_complex}) and $\mathbf{L}$ the Laplacian matrix from
Eq.~(\ref{eq_prob_Lap}). Consequently, the eigenvalues of $\mathbf{G}$ are simply the
eigenvalues of $\mathbf{L}$ divided by the impedance $Z$ and these matrices share the same
eigenvectors.

In this case (the square circuit with identical impedances for its edges), the AC current
flowing between nodes $s$ and $a$ is
$$ \frac{\Delta V_{sa}^{(st)}}{Z_{sa}} = \frac{ I\,e^{j\,\omega\,T} }{Z_{sa}} \left(
    \left[\vec{v}_2\right]_s - \left[\vec{v}_2\right]_a \right) \frac{\left| Z \right|
     }{2\,e^{-j\,\varphi}} \left( \left[\vec{v}_2\right]_s - \left[\vec{v}_2\right]_t \right)\,, $$
\begin{equation}
 I_{sa}^{(st)}(T) = \frac{I}{2}\,e^{j\,\omega\,T} \,,
\end{equation}
which is the same result as in the equal resistances DC case [Eq.~(\ref{eq_current_sol})]
for the modulus because $Z_{sa} = Z$. Furthermore, the analogy is further seen when
calculating the equivalent impedance between the source ($s$) and sink ($t$) nodes using
Eq.~(\ref{eq_equiv_impedance}). This results in
\begin{equation}
  Z_{st}^{(eq)} = \frac{\left| \left[\vec{v}_2\right]_s - \left[\vec{v}_2\right]_t \right|^2
   }{2/Z} = Z\,.
 \label{eq_equiv_sol}
\end{equation}
The solution is identical to the one that Circuit Theory derives and is in direct
correlation with the DC problem as expected.

In more general scenarios, the relationship between the DC and AC circuit is not direct. In
such situations, the complex entries of the Laplacian matrix for the AC case are not related
to the DC Laplacian matrix. Hence, further assumptions need to be done to find analytical
solutions. For instance, one could have to impose that the input frequency to be larger than
the natural frequencies of the lines ($\omega > \omega_{kl}$ for every edge), such that the
imaginary part of the Laplacian be positive semi-defined (see Appendix \ref{Anex_1}).

\section{Conclusions}
The approach we develop provides new analytical insight into the transmission flow problem
and exhibits different features than other available solutions. Moreover, it provides a new
tool to achieve the voltage/current solutions and to analyse resonant behaviour in linear
circuits. As a practical application, we relate these solutions to closed circuits where a
voltage generator is present (instead of having open sources/sinks that feed current to the
network) and solve a simple network where we can compare our solutions to the ones provided
by solving directly Kirchhoff's equations. Our findings help to solve problems, where the
input and output nodes change in time within the network, more effectively than classical
Circuit Theory techniques.

\appendix
\section{Complex weighted Laplacian matrix characteristics}
 \label{Anex_1}
The weighted Laplacian matrix of the circuit network $\mathbf{G}$ with edge properties given
by the symmetric line impedances $Z_{kl} = Z_{lk}$ has the following complex value entries
\begin{equation}
  G_{kl} = \delta_{kl}\left( \sum_{m=1}^M \frac{1}{ Z_{lm} } \right) + \frac{ (\delta_{kl}
   - 1) }{ Z_{kl} }\,,
 \label{eq_Laplacian}
\end{equation}
hence,
\begin{equation}
  \sum_{l = 1}^N G_{kl} = 0\,,\;\forall\,k\,,
 \label{eq_prop1}
\end{equation}
which is the first requirement for a Laplacian matrix: the zero row sum property.

The eigenspace of $\mathbf{G}$ is composed of a set of $N$ complex eigenvalues $\lambda_n$
and eigenvectors $\vec{v}_n$ with $n = 0,1,\ldots,N-1$, such that
$$ \mathbf{G}\vec{v}_n = \lambda_n\vec{v}_n\,,\;\forall\,n\,,$$
thus,
\begin{equation}
  \mathbf{G}\,\mathbf{P} = \mathbf{P}\,\mathbf{\Lambda}\,,
 \label{eq_spectra}
\end{equation}
where $\mathbf{P} = \{ \vec{v}_0,\,\vec{v}_1,\,\ldots,\vec{v}_{N-1} \}$ is a unitary matrix
($\mathbf{I} = \mathbf{P}^{-1} \mathbf{P} = \mathbf{P^\star}^T \mathbf{P}$, $\mathbf{I}$
being the identity matrix) of eigenvectors and $\mathbf{\Lambda}$ is a diagonal matrix of
eigenvalues ($\Lambda_{kl} = \delta_{kl}\,\lambda_{k-1}$).

Due to Eq.~(\ref{eq_prop1}), $\mathbf{G}$ has a null eigenvalue (referred to as $\lambda_0$
in the following) associated to the kernel vector $\vec{v}_0 = \vec{1}/\sqrt{N}$, where
$\vec{1} = (1,\ldots,1)$. Using Eq.~(\ref{eq_prop1}), $\mathbf{G}\vec{1} = \vec{0} =
\lambda_0\vec{1}$. Hence, the kernel of the matrix (the space of eigenvectors associated
to the null eigenvalues) is at least of dimension $1$ and direct inversion of the matrix
is not possible. This is the second property of a Laplacian matrix,
\begin{equation}
 \det(\mathbf{G}) = \det(\mathbf{P}\,\mathbf{\Lambda}\,\mathbf{P}^{-1}) =
  \det(\mathbf{\Lambda}) = \prod_{n=0}^{N-1} \lambda_n = 0\,,
 \label{eq_prop2}
\end{equation}
which implies that the rank of the matrix is less than $N$.

The third property is that Laplacian matrices are positive semi-defined. In particular, for
any column vector $\vec{x}$, the Dirichlet sum is such that
\begin{equation}
  \vec{x}\cdot\mathbf{G}\vec{x} = \frac{1}{2} \sum_{k=1}^M \sum_{l=1}^N W_{kl} \left(x_k -
   x_l\right)^2 \geq 0\,,
 \label{eq_prop3}
\end{equation}
where ``$\cdot$'' is the inner product operation and $W_{kl} = 1/Z_{kl}$ is the weighted
adjacency matrix of the circuit network. This inequality holds only if $W_{kl}\geq0$ for all
$k$ and $l$. As a consequence, it implies that all eigenvalues are non-negative, because
$\vec{x}$ can be any of the $\vec{v}_n$ eigenvectors. In that case,
$$ 0 \leq \vec{v}_n\cdot\mathbf{G}\vec{v}_n = \lambda_n\vec{v}_n \cdot\vec{v}_n = \lambda_n\,,$$
where the last equality is possible because of the unitary property of the eigenvectors
($\vec{v}_n\cdot\vec{v}_m = \sum_{k=1}^N [\vec{v}_n]_k [\vec{v}_m]_k^\star = \delta_{nm}$).
However, Eq.~(\ref{eq_prop3}) is always valid only for Laplacian matrices with non-negative
real entries.

For complex entries, such as in our $\mathbf{G}$, the inequality in Eq.~(\ref{eq_prop3})
can be analysed by splitting the matrix $\mathbf{G}$ into a real ($\mathbf{G}_R$) and an
imaginary ($\mathbf{G}_I$) part, i.e., $\mathbf{G} = \mathbf{G}_R + j\,\mathbf{G}_I$, where
\begin{equation}
 \mathbf{G}_R = \delta_{kl}\left( \sum_{m=1}^M \frac{ \cos\left(\varphi_{lm}\right) }{ \left|
  Z_{lm} \right| } \right) + \frac{ (\delta_{kl} - 1)\cos\left(\varphi_{kl}\right) }{ \left|
   Z_{kl} \right| }\,,
 \label{eq_real_Laplacian}
\end{equation}
\begin{equation}
 \mathbf{G}_I = -\delta_{kl}\left( \sum_{m=1}^M \frac{ \sin\left(\varphi_{lm}\right) }{ \left|
  Z_{lm} \right| } \right) - \frac{ (\delta_{kl} - 1)\sin\left(\varphi_{kl}\right) }{ \left|
   Z_{kl} \right| }\,,
 \label{eq_imag_Laplacian}
\end{equation}
and $Z_{kl} = \left| Z_{kl} \right|\,e^{j\,\varphi_{kl}}$. The Laplacian matrix $\mathbf{G}_R$
contains the information of the network structure resistive part and the Laplacian matrix
$\mathbf{G}_I$ contains the information of the network structure reactive part. In other
words, the dissipative and the resonant structure of the circuit network, respectively.

In order to have a positive (or negative) semi-defined Laplacian matrix, the weighted
adjacency matrix elements need to be positive (or negative) for all pairs of nodes. For
example, for the real part, if $W_{kl}^{(R)} = \cos\left(\varphi_{kl}\right)/\left|Z_{kl}
\right| \geq 0$, then $\mathbf{G}_R$ is positive semi-defined. Consequently, the validity
of this property depends on the magnitude of the phases that the impedance of the lines
introduce. In particular, for lines that can be modelled by series $RLC$, the $W_{kl}^{(R)}
\geq 0$ for every $kl$-edge, hence, $\mathbf{G}_R$ has a non-negative spectra of eigenvalues.
However, in such a case, the sign of $W_{kl}^{(I)} = \sin\left(\varphi_{kl}\right)/\left|
Z_{kl} \right|$ depends on the input/output frequency [see Eq.~(\ref{eq_Z_RLC})]. For $\omega
< \omega_{kl}^{(LC)}$ for all $kl$-edges, $W_{kl}^{(I)} \geq 0$, thus $\mathbf{G}_I$ has a
non-negative spectra of eigenvalues as $\mathbf{G}_R$. For $\omega > \omega_{kl}^{(LC)}$,
$W_{kl}^{(I)} \leq 0$, thus, the opposite happens.

In general, in our case a rule of thumb for knowing the character of the spectra of the
matrix $\mathbf{G}$ is missing (the elements are complex and are not bounded solely to the
positive quadrant of the complex plane). However, the unitary base property of the set of
associated eigenvectors $\{\vec{v}_0,\vec{v}_1,\ldots,\vec{v}_{N-1}\}$ is always valid. This
means that
\begin{equation}
  \vec{v}_n\cdot\vec{v}_m = \sum_{k = 1}^N \left[\vec{v}_n\right]_k \left[\vec{v}_m
   \right]_k^{\star} = \delta_{nm}\,,
 \label{eq_indep}
\end{equation}
and
\begin{equation}
  \left[\mathbf{P}^{-1}\mathbf{P}\right]_{kl} = \sum_{n = 0}^{N-1} \left[\vec{v}_n\right]_k
   \left[\vec{v}_n\right]_l^{\star} = \delta_{kl}\,.
 \label{eq_gen}
\end{equation}

Equation~(\ref{eq_indep}) is the verification that the eigenvector set is composed solely
by linearly independent vectors. Equation~(\ref{eq_gen}) is the completeness property, and
it is the verification that the set is also a generating set. Hence, it conforms a basis of
the linear functions that operate over the set of nodes.

\section{Inversion of the Laplacian matrix and the node Voltage Potential solutions}
 \label{Anex_2}
Due to the existence of the null eigenvalue in any Laplacian matrix, the inverse is ill
defined. We overcome this problem by means of a translation in Eq.~(\ref{eq_model}) and the
removal of the kernel from the eigenvector base. Namely,
\begin{equation}
 \mathbf{G}\,\vec{V}^{(st)} + \frac{\mathbf{J}}{N}\,\vec{V}^{(st)} = \vec{F}^{(st)} +
  \vec{e}^{\,(st)}\,,
 \label{eq_new_model}
\end{equation}
where all entries of matrix $\mathbf{J}$ are equal to unity ($J_{kl} = 1$) and $e_i^{(st)} =
\frac{1}{N} \sum_k V_k^{(st)}$ for all $i = 1,\ldots,N$ is the vector components resultant of
the product between $\mathbf{J}/N$ and $\vec{V}^{(st)}$.

From Eq.~(\ref{eq_spectra}) we know we can write the elements of the Laplacian $\mathbf{G}$
in terms of its complex eigenvalues and eigenvectors by
\begin{equation}
  G_{kl} = \sum_{n = 1}^{N-1} \left[\vec{v}_n\right]_k \lambda_n \left[\vec{v}_n\right]_l^\star,
 \label{eq_spectral_G}
\end{equation}
where the term corresponding to $n = 0$ has been removed because $\lambda_0 = 0$. In a similar
fashion we define the following matrix $\mathbf{X}$ entries
\begin{equation}
  X_{kl} = \sum_{n = 1}^{N-1} \left[\vec{v}_n\right]_k \frac{1}{\lambda_n} \left[\vec{v}_n
   \right]_l^\star.
 \label{eq_spectral_X}
\end{equation}

Here we show that $\mathbf{X} + \mathbf{J}/N$ is the inverse matrix of $\mathbf{G}+\mathbf{J}
/N$. In a sense, $\mathbf{G}$ is a matrix with elements that represent the conductivity of
the edges, whereas $\mathbf{X}$ represents the impedance of the edges. First, we note that
$\mathbf{J}^2 = N\,\mathbf{J}$, hence, $\mathbf{J}^2/N^2 = \mathbf{J}/N$. Then, we observe
that $\mathbf{G}\,\mathbf{J} = \mathbf{0}$ because of the zero row sum property. Similarly,
$\mathbf{X}\,\mathbf{J} = \mathbf{0}$. This is seen from,
$$ \sum_{l = 1}^N X_{kl}\,J_{lm} = \sum_{l = 1}^N X_{kl} = \sum_{n = 1}^{N-1} \left[\vec{v}_n
    \right]_k \frac{1}{\lambda_n} \sum_{l = 1}^N \left[\vec{v}_n\right]_l^\star. $$
However, as Eq.~(\ref{eq_indep}) holds for every eigenvector of $\mathbf{G}$, in
particular, $\vec{v}_0\cdot\vec{v}_m = \vec{1}\cdot\vec{v}_m/\sqrt{N} = \delta_{0m}$,
then $\sum_{l = 1}^N \left[\vec{v}_n\right]_l^\star = 0$ for every spanning eigenvector
($n\neq0$). Finally,
$$ \left[ \mathbf{X}\,\mathbf{G} \right]_{kp} = \sum_{l = 1}^N \left( \sum_{n = 1}^{N-1}
    \left[\vec{v}_n\right]_k \frac{1}{\lambda_n} \left[\vec{v}_n\right]_l^\star
     \sum_{m = 1}^{N-1} \left[\vec{v}_m\right]_l \lambda_m \left[\vec{v}_m\right]_p^\star
      \right) $$
$$ = \sum_{n = 1}^{N-1} \sum_{m = 1}^{N-1} \left[\vec{v}_n\right]_k \frac{1}{\lambda_n}
    \left( \sum_{l = 1}^N \left[\vec{v}_n\right]_l^\star \left[\vec{v}_m\right]_l \right)
     \lambda_m \left[\vec{v}_m\right]_p^\star\,, $$
where, using Eq.~(\ref{eq_indep}), it results in
\begin{equation}
  \left[ \mathbf{X}\,\mathbf{G} \right]_{kp} = \sum_{n = 1}^{N-1} \left[\vec{v}_n\right]_k
    \left[\vec{v}_n\right]_p^\star\,.
 \label{eq_prod}
\end{equation}
Now, observing that Eq.~(\ref{eq_gen}) can be written as
$$ \frac{1}{N} + \sum_{n = 1}^{N-1} \left[\vec{v}_n\right]_k \left[\vec{v}_n\right]_p^{\star}
  = \delta_{kp}\,, $$
then, Eq.~(\ref{eq_prod}) is further simplified
\begin{equation}
  \left[ \mathbf{X}\,\mathbf{G} \right]_{kp} = \delta_{kp} - \frac{1}{N} = \left[\mathbf{I}
   \right]_{kp} - \frac{\left[\mathbf{J}\right]_{kp}}{N}\,.
 \label{eq_prod_XG}
\end{equation}
Consequently, we have shown that
\begin{equation}
  \left( \mathbf{X} + \frac{\mathbf{J}}{N} \right)\left( \mathbf{G} + \frac{\mathbf{J}}{N}
   \right) = \mathbf{I}\,.
 \label{eq_inverse}
\end{equation}

Returning to Eq.~(\ref{eq_new_model}), and using Eq.~(\ref{eq_inverse}), we obtain the
voltage potentials at each node
\begin{equation}
  \vec{V}^{(st)} = \left( \mathbf{X} + \frac{\mathbf{J}}{N} \right)\,\vec{F}^{(st)} +
   \vec{e}^{\,(st)}\,,
 \label{eq_volt_sol}
\end{equation}
where we use that $\mathbf{X}\,\vec{e}^{\,(st)} = \vec{0}$ and $\mathbf{J}\,\vec{e}^{\,(st)}
= N\,\vec{e}^{\,(st)}$. If global conservation of charge is granted, namely, if the input
current equals the output current in phase and magnitude, then, $\mathbf{J}\,\vec{F}^{(st)}
= \vec{0}$. Otherwise, $\mathbf{J}\,\vec{F}^{(st)}$ is a vector with all the elements equal
to the magnitude and/or phase difference between the input and output net currents [see
Eq.~(\ref{eq_netflow})]. We note that the role of $\vec{e}^{\,(st)}$ in Eq.~(\ref{eq_volt_sol})
is to add an arbitrary constant to the node voltage potential. This is easily interpreted as
the arbitrary energy reference point. Such arbitrary value is eliminated once voltage
differences are calculated. Moreover, voltage differences eliminate also the possible
constant value given by $\mathbf{J}\,\vec{F}^{(st)}$. Consequently, the voltage difference
between two arbitrary nodes $k$ and $l$ in the network is given by
\begin{equation}
  \Delta V_{kl}^{(st)} = V_k^{(st)} - V_l^{(st)} = \left[ \mathbf{X}\,\vec{F}^{(st)}
   \right]_k - \left[ \mathbf{X}\,\vec{F}^{(st)} \right]_l\,.
 \label{eq_volt_diff_sol}
\end{equation}
Thus,
$$ \Delta V_{kl}^{(st)} =  F^{in}\left(\, \left[\mathbf{X}\right]_{ks} - \left[\mathbf{X}
    \right]_{ls} \,\right) - F^{out}\left(\, \left[\mathbf{X}\right]_{kt} - \left[\mathbf{X}
     \right]_{lt} \,\right) $$
$$ =  F \left[ \left(\, \left[\mathbf{X}\right]_{ks} - \left[\mathbf{X}
    \right]_{ls} \,\right) - \left(\, \left[\mathbf{X}\right]_{kt} - \left[\mathbf{X}
     \right]_{lt} \,\right) \right] $$
$$ = F\left[ \sum_{n = 1}^{N-1} \left( \left[\vec{v}_n\right]_k - \left[\vec{v}_n\right]_l
    \right)\frac{1}{\lambda_n} \left( \left[\vec{v}_n\right]_s^\star - \left[\vec{v}_n\right]_t^\star
     \right) \right]. $$

\section{Many input/output flows and the relationship to voltage generators}
 \label{Anex_3}
In order to analyse how Eq.~(\ref{eq_volt_diff_sol}) changes when many sources and sinks are
present, we need to rewrite Eq.~(\ref{eq_netflow}) to include the new sources of inflow and
sinks of outflow. Thus, in general, the net current at a node is
\begin{equation}
  \sum_{l = 1}^N I_{kl}^{(\mathcal{V}_s,\mathcal{V}_t)} = F \left( \sum_{s\in\mathcal{V}_s}
   a_s\,\delta_{ks} - \sum_{t\in\mathcal{V}_t} b_t\,\delta_{kt} \right)\,,
 \label{eq_gen_flows}
\end{equation}
where $\mathcal{V}_s$ ($\mathcal{V}_t$) is the set of nodes that act as sources (sinks)
and $a_s$ ($b_s$) is the fraction of the total inflow (outflow) $F$ that goes
through node $s\in\mathcal{V}_s$ ($t\in\mathcal{V}_t$), namely, $\sum_{s\in\mathcal{V}_s}
a_s = 1$ ($\sum_{t\in\mathcal{V}_t} b_t = 1$). Consequently, global conservation
of charge is granted. Substituting Eq.~(\ref{eq_gen_flows}) into Eq.~(\ref{eq_volt_diff_sol}),
the voltage difference between nodes $k$ and $l$ in the circuit network with multiple sources
and sinks is
\begin{eqnarray}
 \nonumber
  \Delta V_{kl}^{(\mathcal{V}_s,\mathcal{V}_t)} = F\left[ \sum_{n = 1}^{N-1} \left(
   \left[\vec{v}_n\right]_k - \left[\vec{v}_n\right]_l\right)\times\frac{1}{\lambda_n}\times
    \right. \\
  \left.\;\;\;\;\;\;\times\left( \sum_{s\in\mathcal{V}_s} a_s\,\left[\vec{v}_n
   \right]_s^\star - \sum_{t\in\mathcal{V}_t} b_t\,\left[\vec{v}_n\right]_t^\star
    \right) \right].
 \label{eq_new_volt_diff_sol}
\end{eqnarray}

\begin{figure}[htbp]
 \begin{center}
  \textbf{(a)}
  \includegraphics[width=1.0\columnwidth]{network_flow_circuit_many_st.eps} \\
  \textbf{(b)}
  \includegraphics[width=1.0\columnwidth]{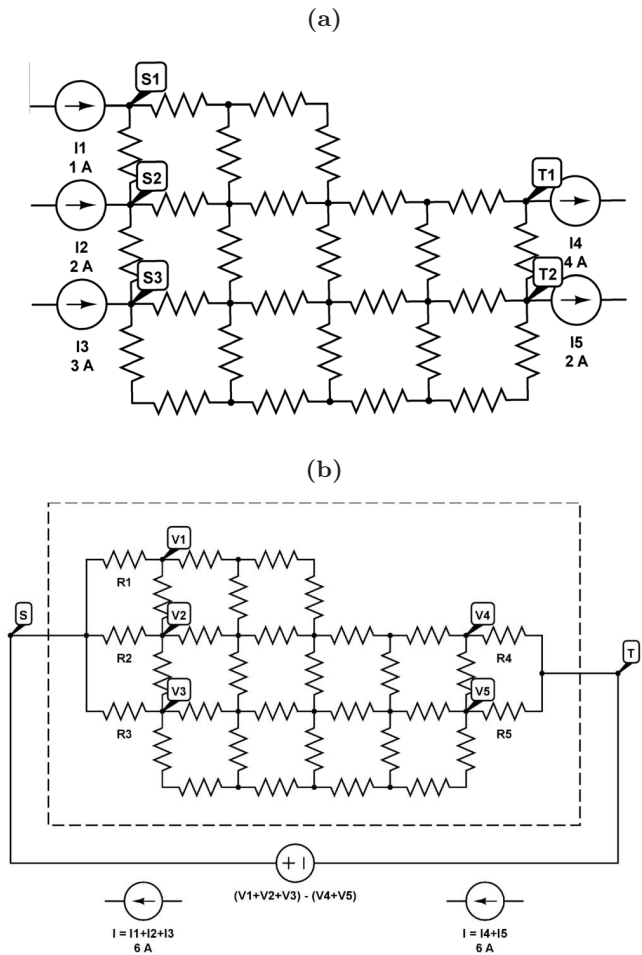}
 \end{center}
  \caption{Panel {\bf (a)} shows a schematic representation of a resistor network with many
  input $I_1,I_2,I_3$ (nodes $S_1,S_2,S_3$) and output $I_4,I_5$ (nodes $T_1,T_2$) currents.
  Panel {\bf (b)} shows the same resistor network but as a closed system containing a
  voltage generator and new resistors. These supply the input (output) currents at nodes
  $S_1,S_2,S_3$ ($T_1,T_2$) via the new resistors ($R_1,R_2,\ldots$) with an identical
  magnitude as in Panel {\bf (a)}.}
 \label{fig_1_anex}
\end{figure}

When multiple sources and sinks exist [e.g., panel {\bf (a)} in Fig.~\ref{fig_1_anex}], then
the transformation of the problem to a closed circuit problem requires the inclusion of a
single super source $S$ node and super sink node $T$ need to be created [panel {\bf (b)}].
All original source (sink) nodes are then connected to the new super source (sink) node that
provides the total input (output) that the multiple sources (sinks) were feeding (consuming)
in the original system $\mathcal{G}$, namely, $I$ ($-I$). Consequently, the multiple
source-sink configuration in $\mathcal{G}$ is transformed into a single $S$-$T$ pair
configuration of a new network $\tilde{\mathcal{G}}$ that has $2$ nodes more than the
original network $\mathcal{G}$. In such conditions, the former process enables to analyse the
new network setting by means of a single generator that connects these two new nodes. In other
words, once a super source $S$ (sink $T$) that connects to all the original sources $S_i\in
\mathcal{V}_s\subset\mathcal{V}$ (sinks $T_j\in\mathcal{V}_t\subset\mathcal{V}$) is defined,
then a Laplace problem can be defined by setting a voltage generator which provides
\begin{equation}
  \tilde{\epsilon}_{ST} = \rho_{ST}\left(\tilde{\mathcal{G}}\right)\,I\,,
 \label{eq_multiple_s-t_transf}
\end{equation}
where the equivalent resistance $\rho_{ST}\left(\tilde{\mathcal{G}}\right)$ between the
super source and super sink is unknown. This is because the impedance (resistance) values
for the new edge connections between the multiple sources $S_i$ (sinks $T_j$) to the super
source [which have to be set such that the current entering the network circuit through the
old multiple sources (sinks) is identical to the one the particular source (sink) supplies
(consumes), e.g., as in panel {\bf (b)} of Fig.~\ref{fig_1_anex}] are unknown.

In order to determine the impedances (resistances) of the edges connecting the super source
(sink) to the multiple source (sink) nodes, we observe that:
$$ \left\lbrace \begin{array}{ll}
        V_S - V_i^{\text{in}} = R_i^{\text{in}}\,a_i\,I\,, & \text{if}\;i\in\{\mathcal{S}\} \\
        V_i^{\text{out}} - V_T = R_i^{\text{out}}\,b_i\,I\,, & \text{if}\;i\in\{\mathcal{T}\}
                \end{array} \right.,$$
where neither the voltages nor the resistances are known. Nevertheless, the voltages of the
super nodes fulfil Eq.~(\ref{eq_multiple_s-t_transf}), thus, arbitrarily setting the unknown
resistances for the new edges to unity, $\rho_{ST}\left(\mathcal{G}^\star\right)$ can be
derived and the node voltages for each of the multiple sources and sinks calculated. That is,
\begin{equation}
  \left\lbrace \begin{array}{ll}
      V_i^{\text{in}}  = V_S - a_i\,I = \rho_{ST} - a_i\,I\,, & \text{if}\;i\in\{\mathcal{S}\} \\
      V_i^{\text{out}} = V_T + b_i\,I = \rho_{ST} + b_i\,I\,, & \text{if}\;i\in\{\mathcal{T}\}
                \end{array} \right..
 \label{eq_multiple_s-t_trick}
\end{equation}

\section*{Acknowledgement}
The authors thank the Scottish University Physics Alliance (SUPA).

\end{document}